\documentclass[aps,prd,twocolumn,notitlepage,superscriptaddress,nofootinbib]{revtex4-1}
\usepackage[utf8]{inputenc}
\usepackage[english]{babel}
\usepackage{amsmath}
\usepackage{amsfonts}
\usepackage{amssymb}
\usepackage{listings}
\usepackage{lipsum}
\usepackage{multirow}
\usepackage{datetime}
\usepackage{graphicx}
\usepackage{mathtools}
\usepackage{mathrsfs}
\usepackage{dcolumn}
\usepackage{multirow}
\usepackage{orcidlink}
\usepackage{color}   
\usepackage{hyperref}
\hypersetup{
    colorlinks=true, 
    pdfborder = {0 0 0.5 [3 3]},
    anchorcolor=black,
    citecolor=blue,
    linktoc=all,    
    linktocpage=true,
    linkcolor=red,
	urlcolor=blue
}

\begin{document}

\title{Ergoregion instability in bosonic stars: \\ scalar mode structure, universality, and weakly nonlinear effects}
\author{Nils Siemonsen\orcidlink{0000-0001-5664-3521}}
\email[]{nils.siemonsen@princeton.edu}
\affiliation{Princeton Gravity Initiative, Princeton University, Princeton NJ 08544, USA}
\affiliation{Department of Physics, Princeton University, Princeton, NJ 08544, USA}

\date{\today}

\begin{abstract} 
Ultracompact spinning horizonless spacetimes with ergoregions are subject to the ergoregion instability. We systematically investigate the instability of a massless scalar field in a variety of rapidly spinning Proca stars and boson stars using WKB-, frequency-, and time-domain methods. We find universal features in the mode structure: the onset of the instability is signaled by a zero-mode, the mode frequencies and growth rates are related by a simple scaling relation in the small-frequency limit (as found for Kerr-like objects), the mode frequencies approach the orbital frequency of counter-rotating stably trapped null geodesics in the eikonal limit, and for each unstable azimuthal mode only a finite number of overtones and polar modes are exponentially growing. The e-folding times are as short as $\tau\sim 10^4 M$ (in terms of the spacetime's ADM mass $M$). Interestingly, we find a near universal relationship between the frequencies and growth rates across all bosonic stars and also compared with Kerr-like objects. Furthermore, we show that weakly nonlinear backreaction of the instability induces a shift in growth rates as well as emission of gravitational waves; we find evidence that these effects lead to an amplification of the unstable process. This suggests that strongly nonlinear interactions are important during the gravitational saturation of the instability.
\end{abstract}

\maketitle

\section{Introduction} \label{sec:Intro}

Sufficiently compact and rapidly spinning spacetimes develop compact ergoregions.  Such regions---where any observer is forced to co-rotate with the spacetime---are ubiquitous in General Relativity and beyond, appearing in a range of solutions from ultracompact horizonless objects~\cite{Cardoso:2019rvt,Bambi:2025wjx,Carballo-Rubio:2025fnc} to black holes and higher-dimensional spacetimes~\cite{Kerr:1963ud,Emparan:2001wn,Emparan:2008eg}. Friedman showed in Ref.~\cite{Friedman:1978} (see also Refs.~\cite{Vilenkin:1978uc,Moschidis:2016zjy}), an asymptotically flat rotating horizonless spacetime with compact ergoregion is linearly unstable to massless perturbations. Concretely, given such a spacetime, then a compact ergoregion is where the asymptotically timelike Killing field $t^\mu$ turns spacelike. The presence of an ergoregion implies the existence of massless states with negative energy defined with respect to $t^\mu$. Such states are time-dependent and radiative, emitting positive energy towards future null infinity. Thus, energy conservation demands that the magnitude of such energy contained in an asymptotically null spacelike foliation grows without bound---the \textit{ergoregion instability} develops. The underlying mechanism of this growth is, in spirit, a Chandrasekhar-Friedman-Schutz (CFS) instability \cite{Chandrasekhar:1970pjp,FS1978a,FS1978b} (for reviews see e.g. Refs.~\cite{Andersson:2000mf,Andersson:2002ch}): the appearance of an ergoregion implies that field perturbations with angular momentum opposite to the spacetime's, which are nominally counter-rotating, are dragged with the spacetime and become co-rotating. While in this frame, the modes rotate with the spacetime, in the spacetime's co-rotating frame, they remain counter-rotating. Under these conditions, the field configurations carry angular momentum of one sign, but emit angular momentum of the opposite sign, leading to an exponential growth at the linear level. Note, this makes the ergoregion instability physically distinct from the superradiance instability of massive fields present in black hole spacetimes \cite{Brito:2015oca}; the former drives negative energy states by emission of positive energy flux at null infinity, while the latter generates positive energy states by absorption of negative energy flux at the horizon. 

In the context of axisymmetric compact objects, the azimuthal dependence of a linear massless bosonic field is $e^{im\varphi}$, where $m$ is the azimuthal mode number. In the eikonal limit, $m\rightarrow \infty$, in which the dynamics approach that of null geodesics trapped inside the ergoregion, any such mode must co-rotate with the spacetime. Hence, the above mechanism implies, for sufficiently large $m_0>0$, there exist field modes with $m>m_0$, which grow exponentially quickly \cite{Friedman:1978}. In the eikonal limit, the e-folding time $\tau$ of the ergoregion instability grows exponentially with mode number \cite{Comins:1978}: $\tau\sim \exp(\beta m)$, with $\beta>0$, for $m>m_0$. For instance, within uniform density fluid stars of mass $M$ the $m\geq m_0= 2$ scalar modes are unstable with timescales\footnote{We set $G=c=1$ here, and in the following.} $\tau/M\gtrsim \mathcal{O}(10^7)$ \cite{Comins:1978, Yoshida:1996}. Considering gravitational perturbations of Kerr-like objects the timescales can be as short as $\tau/M \gtrsim \mathcal{O}(10^4)$ (see e.g., Ref.~\cite{Maggio:2017ivp}). Beyond this, the ergoregion instability appears in a variety of other contexts from neutron stars to string theory and gravitational wave astronomy~\cite{Schutz:1978, Kokkotas:1992, Kokkotas:2002sf,Cardoso:2007az, Chirenti:2008pf, Cardoso:2005gj, Pani:2010jz, Oliveira:2014oja,Maggio:2017ivp, Vicente:2018mxl,Cardoso:2008kj,Barausse:2018vdb,Tsokaros:2019mlz,Tsokaros:2020qju, Ruiz:2020zaz, Dey:2020pth,Chakravarti:2023wlc,Bianchi:2023rlt,Oliveira:2024quw,Zhong:2022jke,Franzin:2022iai,filipe2025,Saketh:2024ojw,Chowdhury:2007jx} (see also related work in Refs.~\cite{Keir:2018hnv,Eperon:2016cdd,Keir:2016azt}). This test field instability differs from unstable linear perturbations of non-vacuum solutions to the Einstein equations with ergoregions, where matter and gravitational degrees of freedom are coupled, such as unstable $w$-modes in fluid stars~\cite{Kokkotas:2002sf}. Furthermore, while dissipative effects can quench the instability \cite{Ips1991,Lindblom:1998wf,Andersson:1998ze,Maggio:2018ivz}, it may also enable other types of growth mechanisms \cite{Rob1963,Ski1996,Andersson:2002ch,Bonazzola:1995zu,Redondo-Yuste:2025ktt}, and radiative dissipation may even serve to amplify the ergoregion instability. Lastly, the nonlinear evolution of the instability is necessarily connected to the nonlinear dynamical properties of the stable light ring (and the conjectured nonlinear light ring instability \cite{Keir:2014oka,Cardoso:2014sna}), as demonstrated explicitly in Ref.~\cite{Siemonsen:2025fne,Siemonsen:2025ucx}. Based on the dynamics of the ergoregion instability, it has been argued that gravitational wave observations can be used to place constraints on the existence of ultracompact and black hole mimicking objects \cite{Fan:2017cfw,Barausse:2018vdb,Mastrogiovanni:2025ixe,Maggio:2021ans} (see also Ref.~\cite{Cardoso:2019rvt}). 

In this work, we study the ergoregion instability of a massless scalar test field propagating on spinning bosonic star spacetimes and discuss leading backreaction effects. With that, we lay the groundwork for a series of forthcoming works on the fully \textit{nonlinear} development of the ergoregion instability in these spacetimes. These horizonless and ultracompact objects, sourced by massive scalar \cite{PhysRev.172.1331, PhysRev.187.1767} or vector \cite{Brito:2015pxa} fields, serve as proxies for a larger set of black hole mimickers. In particular, bosonic stars are useful, because (i) they are currently the only such ultracompact object, which can be treated within established numerical relativity frameworks \cite{Liebling:2012fv}, and (ii) in contrast to almost all other ultracompact objects, non-perturbatively spinning boson star solutions (with ergoregions) have been found \cite{Yoshida:1997qf,Kleihaus:2005me,Kleihaus:2007vk}. The ergoregion instability of a massless scalar field on scalar boson star backgrounds was investigated in Ref.~\cite{Cardoso:2007az}, though, with very limited scope. To understand the ergoregion instability at the linear level systematically, we proceed in this work by analyzing a series of both scalar and vector bosonic stars, in the following referred to simply as boson stars (BSs) and Proca stars (PSs), respectively. All these spacetimes are stationary, axisymmetric, asymptotically flat, horizonless, and everywhere regular; our focus lies on those rapidly spinning stars with compact ergoregion.

We begin our analysis by comparing three distinct methods to compute the ergoregion unstable scalar test field modes in these spacetimes: a WKB method, an approximate frequency-domain method, and a complete time-domain method. While the frequency-domain method captures all relevant qualitative features of the instability, it can differ significantly from the time-domain results as it makes a slow-rotation assumption; the WKB method generally over-predicts the instability growth rate across the parameter space by orders of magnitude, particularly for small azimuthal indices. The time-domain method, on the other hand, is the most accurate, but computationally costly. Therefore, utilizing the frequency-domain method, we find the following general properties of the unstable scalar mode structure: 
\begin{itemize}
\item[(i)] the onset of the ergoregion instability is signaled by a zero-frequency mode.
\item[(ii)] if the zeroth overtone of the $\ell=m$ mode is unstable, then we find there exists only a finite number of accompanying unstable polar modes with $\ell_0\geq\ell\geq m$ and radial overtones with $n_0\geq n\geq 0$.
\item[(iii)] the frequencies $\omega_R$ and growth rates $\omega_I$ of the $\ell=m$ modes are related by $\omega_I M\sim |\omega_R M|^{2\ell+1}$ in the small-frequency limit.
\item[(iv)] in the eikonal limit $\ell=m\rightarrow\infty$, the unstable mode's frequencies approach $\ell \omega_-$, where $\omega_-$ is the orbital frequency of null geodesics in the equatorial counter-rotating stable light ring.
\item[(v)] the instability growth rates with $\omega_IM\lesssim 10^{-5}$ on bosonic star backgrounds are comparable to, or surpass those of, Kerr-like objects.
\end{itemize}
Interestingly, we find a near-universal relationship between growth rates and frequencies across all considered families of bosonic stars and also compared with the corresponding unstable modes in Kerr-like objects.

Additionally, we discuss weakly nonlinear gravitational effects that may become relevant as the unstable field acquires significant energy and angular momentum during the late stages of the unstable growth. In particular, a shift of the unstable frequency and growth rate as well as gravitational wave emission are able to either accelerate or weaken the growth. In the case of a spinning PS, we find evidence that the growth rate \textit{increases} in the weakly nonlinear regime, thereby \textit{enhancing} the unstable process and suggesting that strongly nonlinear effects will be important in dictating the final fate of the instability. We also find that the additional dissipation channel of gravitational waves remains subdominant before strongly nonlinear effects become important.

\section{Methods}

We begin by discussing and summarizing basic properties of the families of bosonic star solutions considered in this work, serving as the background spacetimes for the scalar test field. Following this, we introduce the three methods (WKB, frequency-domain, and time-domain) to compute unstable massless scalar test field configurations, and in particular their frequencies and growth rates. We conclude by comparing these methods and outline their range of validity. 

\subsection{Stationary bosonic stars} \label{sec:statBS}

Bosonic stars are stationary asymptotically flat axisymmetric solutions supported by the self-gravity of the complex scalar and vector field sourcing the spacetime. As such, they are solutions in the theory
\begin{align}
\begin{aligned}
S=\int d^4x\sqrt{-g}\bigg[ \frac{R}{16\pi} & \ -g^{\alpha\beta}\nabla_{(\alpha} \bar{\Psi}\nabla_{\beta)} \Psi-V(|\Psi|) \\
 & \ -\frac{1}{4}F_{\alpha\beta}\bar{F}^{\alpha\beta}-\frac{\mu_A^2}{2} A_\alpha \bar{A}^\alpha \bigg].
\label{eq:action}
\end{aligned}
\end{align}
Here $V(|\Psi|)$ is the scalar potential, which necessarily contains the mass-term $\mu^2_S|\Psi|^2$, of the complex scalar field $\Psi$, $R$ is the Ricci scalar associated with the metric $g_{\mu\nu}$, and $A_\mu$ is the complex vector field of mass $\mu_A$ and associated field strength $F_{\mu\nu}$. In the following we drop the subscript on the mass parameters for brevity and overbars denote complex conjugation. Assuming the above symmetries, the metric can be written in Lewis-Papapetrou form
\begin{align}
\begin{aligned}
ds^2=-f dt^2+l f^{-1}\big\{ & g(dr^2+ r^2 d\theta^2) \\
+ & r^2\sin^2\theta \left( d\varphi - \Omega r^{-1} dt\right)^2 \big\},
\label{eq:metricansatz}
\end{aligned}
\end{align}
with the set of functions $\{f,l,g,\Omega\}$ depending on both $r$ and $\theta$, and we choose $\Omega<0$. In general, bosonic star solutions to the equations following from \eqref{eq:action} can be obtained only numerically; throughout this work, we utilize construction methods outlined in Refs.~\cite{Siemonsen:2020hcg,Siemonsen:2023hko} (see also Ref.~\cite{Kleihaus:2005me}).

\begin{figure*}[t]
\includegraphics[width=1\textwidth]{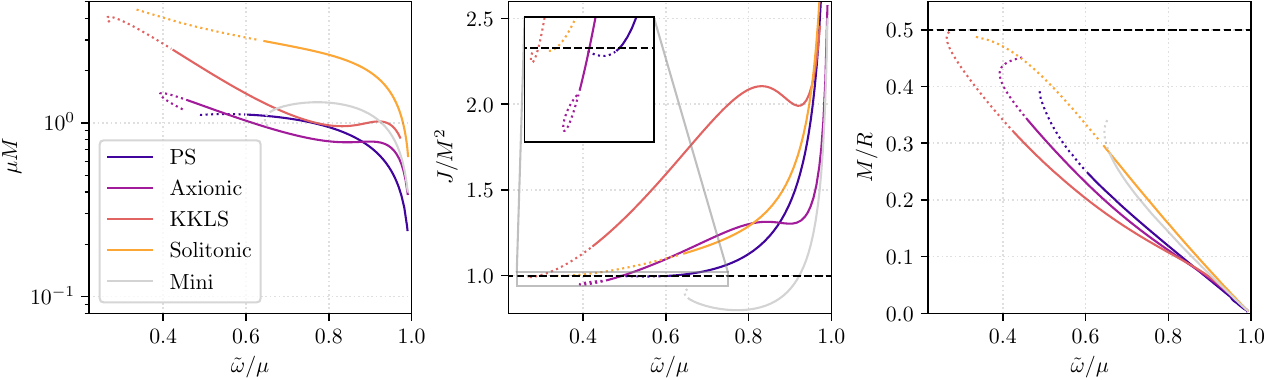}
\caption{The mass $M$, the angular momentum $J$, and the compactness $C=M/R$ of the families of bosonic star solutions relevant for this work. Each is parametrized by the internal frequency $\tilde{\omega}$. Specifically, we focus on the $\tilde{m}=1$ axionic family with $\tilde{f}=0.025$, the $\tilde{m}=2$ KKLS family with $\kappa=0.1$, the $\tilde{m}=3$ solitonic family with $\tilde{\sigma}=0.2$, and the $\tilde{m}=1$ PS set of solutions; for comparison, we also show the properties of the $\tilde{m}=1$ mini BS family (``Mini'' refers to solutions in a scalar model without self-interactions; e.g., $f,\tilde{\sigma}\rightarrow\infty$). The $\tilde{\omega}/\mu\approx 1$ regime is the Newtonian limit, where $C\rightarrow 0$; a non-rotating black hole has a compactness of $C=1/2$. Dotted lines indicate those solutions along each family, which exibit ergoregions. Notice, the dimensionless spin approaches a value below unity only in the highly relativistic limit, as the compactness of the stars approaches that of a black hole.}
\label{fig:family}
\end{figure*}

To use these solutions as test-beds for the evolution of the ergoregion instability, we are interested in highly-compact stars with large ergoregions. In the case of BSs, such solutions exist only in models with significant self-interactions. Families of scalar star solutions in these models typically develop a second (more compact) branch of solutions with ergoregions in large portions of the parameter space below their maximum mass \cite{Kleihaus:2007vk, Kleihaus:2011sx}. Additionally, in an attempt to study a diverse set of such solutions, we consider a series of different scalar self-interactions leading to star solutions with a variety of global properties. Explicitly, we consider the solitonic potential \cite{Friedberg:1986tq}
\begin{align}
V(|\Psi|)=\mu^2|\Psi|^2\left(1-\frac{2|\Psi|^2}{\tilde{\sigma}^2}\right)^2,
\label{eq:solitonic}
\end{align}
with coupling $\tilde{\sigma}$, the axionic potential \cite{Siemonsen:2020hcg}
\begin{align}
V(|\Psi|)=\mu^2 \tilde{f}^2\big\{1-\cos[\sqrt{2 |\Psi|^2}\tilde{f}^{-1}]\big\}.
\label{eq:axionpot}
\end{align}
with coupling $\tilde{f}$, and finally, the KKLS potential \cite{Kleihaus:2005me, Kleihaus:2007vk, Kleihaus:2011sx},
\begin{align}
V(|\Psi|)=\mu^2|\Psi|^2\left[1 -\frac{16\pi}{1.1 \kappa}|\Psi|^2+ \frac{64\pi^2}{1.1\kappa^2} |\Psi|^4\right],
\label{eq:metastable}
\end{align}
with coupling $\kappa$. PSs in the minimal model \eqref{eq:action} exhibit ergoregions without the need for self-interactions.

Bosonic stars are characterized by their harmonic azimuthal number $\tilde{m}$ and frequency $\tilde{\omega}$, i.e., $\Psi\sim e^{i(\tilde{\omega} t+\tilde{m}\varphi)}$ and $A_\mu\sim e^{i(\tilde{\omega} t+\tilde{m}\varphi)}$, respectively. In particular, for each index $\tilde{m}$ there exists a \textit{one}-parameter family of solutions, parametrized by, for instance, $\tilde{\omega}$.\footnote{We mostly restrict to the branch of solutions below the maximum mass and connected to the non-relativistic limit; there this parameterization is unique. We also focus entirely on those solutions in their radial ground states.} The mass $M$ and angular momentum $J$ of a given representative of a family of solutions are obtained from the Komar expressions (see e.g., Ref.~\cite{Siemonsen:2020hcg} for details), while the radius $R$ is defined as the circular radius inside which 99\% of the mass lies. Throughout this work, we restrict to the following four one-parameter families of rotating bosonic star solutions: the $\tilde{m}=1$ family of stars in the axionic model with coupling $\tilde{f}=0.025$ (labeled simply as ``Axionic''), the $\tilde{m}=2$ KKLS family with parameter $\kappa=0.1$ (labeled ``KKLS''), the $\tilde{m}=3$ solitonic set of solutions with scalar coupling $\tilde{\sigma}=0.2$ (referred to simply as ``Solitonic''), and finally, the $\tilde{m}=1$ family of PSs. The main properties of these solutions are summarized in Fig.~\ref{fig:family}. Solutions beyond the maximum mass of each family are expected to be linearly unstable. Except for the PS and ``Axionic'' families, most shown solutions are likely unstable to a non-axisymmetric instability (not related to the ergoregion) \cite{Sanchis-Gual:2019ljs,Siemonsen:2020hcg}, even below the maximum mass of the family of solutions.

\subsection{Integration methods} \label{sec:testfieldmethods}

We are interested in the behavior of a minimally coupled massless scalar test field, and its ergoregion instability, propagating on the bosonic star background spacetimes introduced in the previous section. The underlying field equation is simply the Klein-Gordon equation
\begin{align}
\square_g \Phi=0,
\label{eq:KGequation}
\end{align}
on these fixed spacetimes. Due to the spacetime's symmetries, the field $\Phi$ takes on the form $\Phi\sim e^{-i\omega t}e^{im\varphi}$, where the non-vanishing imaginary component of $\omega=\omega_R+i\omega_I$ reflects the dissipative nature of these asymptotically flat settings. To understand the ergoregion instability of this minimal model, schemes are needed to solve \eqref{eq:KGequation} for the relevant unstable field configurations. The primary difficulty with solving \eqref{eq:KGequation}, given appropriate boundary conditions, lies with the background spacetime's angular momentum. Standard scalar field ansätze using spherical harmonics fail to separate the radial and polar sectors of the partial differential equation outside the slow-rotation limit. For example, in the Kerr spacetime an alternate sequence of functions exists, spheroidal harmonics, which enable the complete separation of \eqref{eq:KGequation} for arbitrary spins \cite{Teu1973,Pre1973}. For bosonic stars, no such sequence is known. To tackle this problem, here we utilize two existing approximate semi-analytic techniques, which have been applied to a set of ergoregion unstable spacetimes, and contrast these with fully numerical time-domain evolutions of \eqref{eq:KGequation} from appropriately chosen initial data. The first approximation technique is a frequency-domain method, assuming a slow-rotation form of the metric \cite{Hartle1967} with
\begin{itemize}
	\item[(i)] $l(r,\theta)\rightarrow l_e(r):= l(r,\theta)|_{\theta=\pi/2}$, etc., and $g\equiv 1$.
\end{itemize}
Secondly, a WKB method, which makes use of (i) and is valid only in the eikonal limit, i.e., assumes
\begin{itemize}
	\item[(ii)] the azimuthal index is large: $m\rightarrow\infty$.
\end{itemize}
Generally, these two methods can efficiently be applied to most of the relevant parameter space, at reduced accuracy. On the other hand, fully numerical time-domain methods, making neither of these assumptions, are providing the most accurate solutions and are applicable only in settings with relatively short instability timescales. In the following, we introduce all three methods in more detail.

\subsubsection{Direct integration method} \label{sec:di_method}

Beginning with the direct integration method, i.e., making assumption (i), the scalar field ansatz based on \textit{spherical} harmonics, $Y_{\ell m}(\theta,\varphi)$,
\begin{align}
\Phi=\sum_{\ell\geq 0}\sum_{|m|\leq \ell} \int d\omega \chi_{\ell m}(r)Y_{\ell m}(\theta,\varphi )e^{-i\omega t},
\label{eq:ScalarfieldAnsatz}
\end{align}
reduces the problem to solving an ordinary differential eigenvalue problem for $\chi_{\ell m}(r)$, with eigenvalue $\omega$, given appropriate boundary conditions. In the Lewis-Papapetrou form of the metric \eqref{eq:metricansatz}, together with assumption (i), the massless Klein-Gordon equation \eqref{eq:KGequation} with ansatz \eqref{eq:ScalarfieldAnsatz} is equivalent to the radial equation
\begin{align}
\begin{aligned}
\frac{1}{r^2\sqrt{l_e}}\frac{d}{dr}\bigg[r^2\sqrt{l_e} & \frac{d\chi_{\ell m}}{dr}\bigg] \\ 
-\bigg[ \frac{\ell(\ell+1)}{r^2} &-\frac{l_e(\omega r -m \Omega_e)^2}{f_e^2r^2}\bigg]\chi_{\ell m}=0.
\end{aligned}
\label{eq:radscaleq}
\end{align}
Unstable field configurations associated with the ergoregion instability, exhibit outgoing flux at null infinity and are regular at the center of the bosonic stars. In this context, the purely outgoing radiation condition is
\begin{align}
\lim_{r\rightarrow\infty}\chi_{\ell m}\sim r^{-1}e^{i\omega r}.
\label{eq:outgoingcondition}
\end{align}
Provided these boundary conditions, we numerically integrate \eqref{eq:radscaleq} radially inwards starting deep in the radiation zone $r\gg M$, using standard Runge-Kutta methods. The eigenvalue $\omega$ is obtained by minimizing $|\chi_{\ell m}(0)|^2$ iteratively.\footnote{The radial equation \eqref{eq:radscaleq} has a regular singular point at the origin. Hence, one can also expand the solution in a Frobenius series as $\chi_{\ell m}\sim r^\ell(1+c_2r^2+ \dots)$, with $c_2=-\omega^2 l_e(0)[2f_e(0)^2(3+2\ell)]^{-1}$, and integrate radially outwards.} Due to the nature of the instability, not all $m$-modes are unstable; this is evident from the reflection symmetry, $m\rightarrow -m$ and $\omega\rightarrow -\omega$, in \eqref{eq:radscaleq}, which switches the sign of the imaginary part $\omega_I$. Therefore, in the following we focus entirely on the unstable modes, which here correspond to the $m>0$ modes (since we chose $\Omega<0$). For each tuple $(\ell,m)$ there exist a series of solutions with a discrete number of radial nodes $n$; we label these the overtones of the fundamental (i.e., $n=0$) solution. In the following, we refer to frequencies and growth rates of unstable modes obtained using this approach as the \textit{direct integration estimates}.

\begin{figure}[t]
\includegraphics[width=0.49\textwidth]{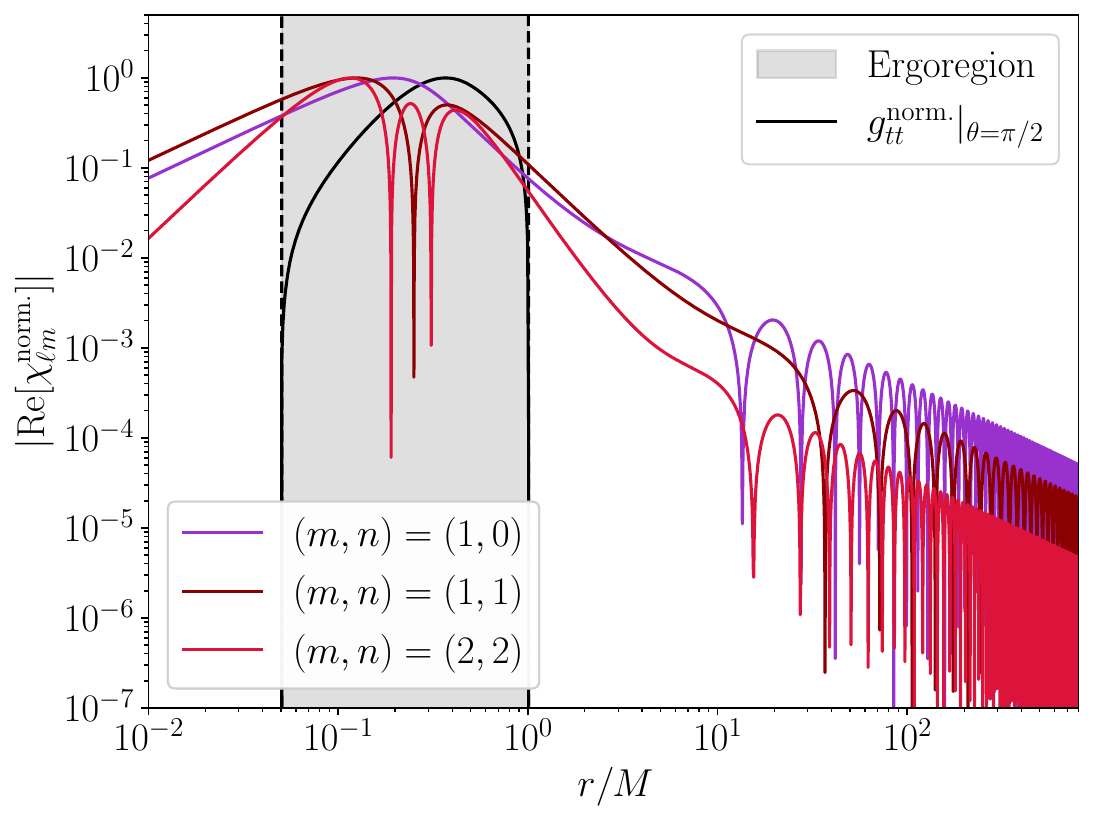}
\caption{The absolute value of the real part of three solutions to \eqref{eq:radscaleq} with maximum normalized to unity obtained with the direction integration method. Here we show only $\ell=m$ modes of varying overtone number $n$. For reference, we show also the maximum-noramlized metric component, $g_{tt}^\text{norm.}$, indicating the location of the ergoregion.} 
\label{fig:chilm}
\end{figure}

In \figurename{ \ref{fig:chilm}}, we present examples of the solutions of \eqref{eq:radscaleq} for a set of azimuthal $\ell=m$ and overtone $n$ numbers of the scalar field propagating in the $\tilde{m}=2$ KKLS BS spacetime with frequency $\tilde{\omega}/\mu=0.281$. The solution reproduces the asymptotic behavior $\sim r^\ell$ at the origin and approaches an outwards radiating $\sim r^{-1}$ solution at large distance, as imposed by \eqref{eq:outgoingcondition}. The solution reaches its global maximum inside the ergoregion and decays rapidly across the ergosurface. 

\subsubsection{WKB method}

The ergoregion instability is active in the large-$m$ regime of modes of the form \eqref{eq:ScalarfieldAnsatz} \cite{Friedman:1978}. For this reason, Comins and Schutz used the WKB approximation in Ref.~\cite{Comins:1978} to determine both the unstable mode's frequencies and growth timescales $\tau=\omega_I^{-1}$ (see also Ref.~\cite{Cardoso:2007az}). This approach not only makes assumption (i), but its validity is also restricted to the eikonal limit, i.e., satisfying assumption (ii). Following Ref.~\cite{Comins:1978}, we restrict to the $\ell=m$ family of modes, redefine the radial functions as $\chi_{\ell=m,m}=\exp[-\int dr(2/r+l'_e/l_e)/2]\tilde{\chi}_{m}$, and apply assumption (ii). Then the radial problem \eqref{eq:radscaleq} simplifies further to \cite{Comins:1978}
\begin{align}
\frac{1}{m^2}\frac{d^2\tilde{\chi}_m}{dr^2}=-T_{m\Sigma}\tilde{\chi}_m, & & \Sigma=\frac{\omega_R}{m},
\label{eq:WKBradeq}
\end{align}
where
\begin{align}
\begin{aligned}
T_{m\Sigma}=\frac{l_e}{f^2_e}(\Sigma-V_+)(\Sigma-V_-), & & V_\pm=\frac{\Omega_e}{r}\pm\frac{f_e}{r\sqrt{l_e}}.
\end{aligned}
\label{eq:eikonalpot}
\end{align}
The WKB approximation can be directly applied to the above radial equation together with potential $T_{m\Sigma}$ to determine the eigenfrequencies and growth timescales in this eikonal limit. To that end, the radial domain is split into four regimes: The inner region, $(0,r_<)$, where $V_+>\Sigma$, the classically allowed region $(r_<,r_>)$, where $V_+<\Sigma$, the tunneling region, $(r_>,r_-)$, where $V_+>\Sigma$ and $V_-<\Sigma$, and the radiation region, $(r_-,\infty)$, where $V_->\Sigma$. The unstable modes' frequencies with overtone number $n$ are then determined by \cite{Comins:1978}
\begin{align}
m\int_{r_<}^{r_>}dr\sqrt{T_{m\Sigma}}=\pi \left(n+\frac{1}{2}\right),
\label{eq:WKBfreq}
\end{align}
while the associated timescales are given by
\begin{align}
\tau=4\exp\left[2m\int_{r_>}^{r_-}dr \sqrt{T_{m\Sigma}}\right]\int_{r_<}^{r_>}dr \frac{d\sqrt{T_{m\Sigma}}}{d\Sigma}.
\label{eq:WKBtimescales}
\end{align}
In the following, we refer to modes and timescales obtained through \eqref{eq:WKBfreq} and \eqref{eq:WKBtimescales}, which make assumptions (i) and (ii), as the \textit{WKB estimates}. 

\subsubsection{Time-domain method} \label{sec:td_method1}

While dismissing assumption \textit{(ii)} not necessarily increases the complexity of the problem significantly, dropping \textit{(i)} requires potentially solving a partial differential equation, in the frequency-domain. This lies beyond the scope of this work. However, we can circumvent these difficulties by solving the system in the time-domain. In that setting, all assumptions above are dropped naturally. In the time-domain, the massless scalar field equation \eqref{eq:KGequation} is integrated forward in time, starting from appropriate scalar initial data. The background geometry is fixed by the chosen bosonic star; we return to relaxing this assumption to study weakly nonlinear effects in Sec.~\ref{sec:nonlinear}. The scalar field solutions obtained in the direct integration approach pose as excellent initial data to trigger the ergoregion instability in the time-domain. Let $S\subset\Sigma_t$ be a compact subset of the $t=\text{const.}$ Cauchy-slice in coordinates \eqref{eq:metricansatz}, $\xi^\mu$ the asymptotically timelike Killing field associated with the background's stationarity, and the scalar stress energy tensor
\begin{align}
T_{\mu\nu}=2\partial_{(\mu} \bar{\Phi}\partial_{\nu)}\Phi- g_{\mu\nu} g^{\alpha\beta}\partial_{(\alpha}\bar{\Phi}\partial_{\beta)}\Phi.
\label{eq:setensor}
\end{align}
The fastest growing field configurations are identified by their \textit{positive} scalar energy flux at large distance, $\dot{E}_\Phi$ defined on $\partial S$, and \textit{negative} energy functional
\begin{align}
E_\Phi=-\int_S d^3x\sqrt{-g} T^t{}_\nu \xi^\nu.
\label{eq:energyfunctional}
\end{align}
To solve \eqref{eq:KGequation} numerically in the time-domain, we proceed as described in Ref.~\cite{Siemonsen:2020hcg} and make use of the generalized Cartoon method \cite{Pretorius:2004jg}. This enables us to fix the azimuthal index $m$ of the scalar field exactly throughout the evolutions. The non-vanishing spacetime angular momentum dictates that the ergoregion unstable states have support over several polar spherical harmonic $\ell$-modes at fixed $m$. Therefore, given the initial data of the form \eqref{eq:ScalarfieldAnsatz}, during an initial transient the field populates these modes and settles into the fastest growing field configuration fully consistently satisfying \eqref{eq:KGequation} at late times. In Appendix~\ref{app:NumSetup}, we discuss the transient and the limitations of this time-domain approach in more detail. Here we note simply, despite the rapidly spinning background spacetimes, with $J\approx M^2$, the spherical harmonic components of the most unstable modes exhibit exponentially small overlap with neighboring $\ell$-modes. Given these time-domain solutions, we extract the frequency by computing the flux ratio $\omega_R\approx m\dot{E}_\Phi/\dot{J}_\Phi$ of the scalar energy flux $\dot{E}_\Phi$ to its angular momentum flux $\dot{J}_\Phi$ on $\partial S$. The growth timescale is obtained from an exponential fit to the total energy $-E_\Phi$ and compared with $\omega_I\approx -\dot{E}_\Phi/(2E_\Phi)$. All unstable test field frequencies and timescales obtained with this time-domain method are referred to as \textit{time-domain estimates} in the following.

\subsection{Method comparison}
\label{sec:methodcomparison}

\begin{figure}[t]
\includegraphics[width=0.485\textwidth]{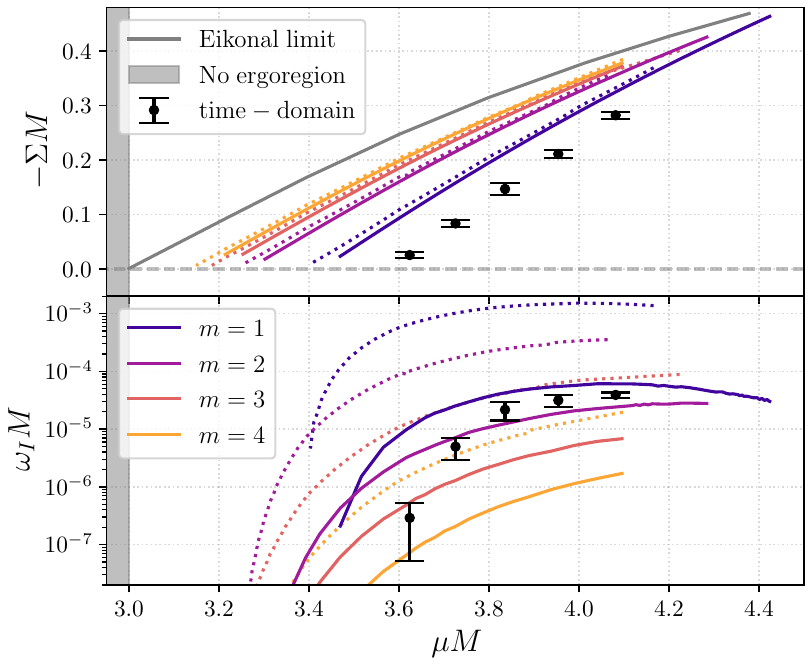}
\caption{Estimates for the ergoregion instability growth rates $\omega_I$ and frequencies $\Sigma=\omega_R/m$ of a series of $\ell=m$ and $n=0$ unstable modes along the sequence of $\tilde{m}=3$ solitonic stars parametrized by their total mass $M$. The dotted lines correspond to the WKB estimates, the solid lines to those obtained using the direct integration method, and the data points indicate the results obtained using our time-domain methods for the $m=1$ mode (a discussion of the error bars can be found in Appendix~\ref{app:NumSetup}). Background star solutions with mass $\mu M\lesssim 3$ exhibit no ergoregions, while the (coordinate) size of the ergoregion grows with $M$ above this critical value. We also show the eikonal limit of the frequencies $\Sigma$; a discussion of this is deferred to Sec.~\ref{sec:eikonal}.}
\label{fig:testfieldcomp}
\end{figure}

Having introduced three different methods to compute ergoregion unstable mode frequencies and growth rates, we are now able to assess the accuracy of each in the context of highly compact and relativistic bosonic stars. To that end, we focus on the family of $\tilde{m}=3$ solitonic stars. Consulting Fig.~\ref{fig:family}, above a critical mass of $\mu M\approx 3$, this family of solutions exhibits an ergoregion, whose (coordinate) size increases with increasing $M$. In \figurename{ \ref{fig:testfieldcomp}}, we compare both the frequency, $\Sigma=\omega_R/m$, and growth rate, $\omega_I$, estimates obtained by each method along this sequence of spacetimes. 

Let us first compare the WKB and direct integration methods. The frequencies of the unstable modes agree to within $\mathcal{O}(10\%)$ even for $m=1$. Naturally, towards the eikonal limit the disagreement decreases. The WKB method generically underestimates the size of the ergoregion at which a given $m$-mode turns unstable; i.e., the zero-crossing, where $-\omega_R,\omega_I\rightarrow 0$, occurs at smaller $M$ (and hence, smaller ergoregions). The growth rates, on the other hand, can differ by several orders of magnitude. This disagreement is driven by two effects. Close to the onset of the instability, the disagreement simply originates from slightly different zero-crossings of the modes, while in the regime with $-\Sigma M\sim\mathcal{O}(0.1)$, the direct integration method is expected to be more accurate away from the eikonal limit. In Fig.~\ref{fig:testfieldcomp}, comparing the growth rates at $\mu M = 4$, the difference between the WKB and direct integration estimates decreases with increasing $m$. Hence, for rapidly spinning bosonic stars the growth rates of the WKB approximation are valid at best at the order-of-magnitude level.

Moving up in accuracy, lets compare the direct integration estimates with the time-domain estimates. As can be seen from Fig.~\ref{fig:testfieldcomp}, the difference in frequency $\omega_R$ between these methods can be above the $\sim 10\%$ level, particularly, close to the zero-crossings. In direct analogy to the previous comparison, the difference in the growth rate is largest close to those crossings due to the slight shift in the onset of the instability. In the $-\Sigma M\sim\mathcal{O}(0.1)$ region of the parameter space, the direct integration and time-domain growth rate estimates are in much better agreement. For instance, at $\mu M\approx 3.96$ these estimates agree to within less than a factor of two. Since the time-domain method is our most accurate, we simply quote numerical uncertainties in Fig.~\ref{fig:testfieldcomp} to estimate its precision.

\section{Linear instability results}

The spectrum of unstable modes of a massless scalar test field exhibits various different features. Most of these features are likely not specific to bosonic stars, but rather universal and appear in a larger class of horizonless spacetimes with ergoregions. In the following, we explore these properties one by one across the parameter space of bosonic star models. To that end, the direct integration method introduced in the previous section is most suited as it is readily applied to a large set of spacetimes, while providing more accurate results than the WKB approximation. In this section, we focus entirely on scalar test fields and turn to a discussion of weakly nonlinear effects in Sec.~\ref{sec:nonlinear}.

\subsection{Zero-mode}
\label{sec:zeromode}

The onset of the ergoregion instability along a sequence of background spacetimes is marked by a stationary zero-frequency mode, as expected within the context of the CFS mechanism \cite{friedman1975a} (see e.g., Ref.~\cite{Andersson:2000mf}). This can be understood as follows: let $\lambda$ parametrize the sequence of horizonless spacetimes with compact ergoregion, $\omega(\lambda)=\omega_R(\lambda)+i\omega_I(\lambda)$ be the frequency of a given mode solution associated with a set of indices\footnote{In this work, the azimuthal and polar index, as well as the overtone number.}, and $\lambda=0$ identify the onset of the instability, i.e., $\omega_I(0)=0$ with $\omega_I<0$ for $\lambda<0$. In this context, the energy, $E$, and angular momentum, $J$, of the mode are related by $\omega_R=mE/J$. Those scalar field solutions with outgoing radiation that turn unstable are those with positive angular momentum\footnote{Recall, the background spacetime's angular momentum is negative in our convention, since $\Omega<0$.} $J>0$ (for any $\lambda$). The azimuthal index $m$ remains fixed, while the energy changes sign along the family of solutions, i.e., $E(\lambda<0)>0$ and $E(\lambda>0)<0$ \cite{Friedman:1978}. This implies that $\omega_R(0)=mE/J|_{\lambda=0}=0$ at the onset of the instability. Notice, while the energy flux at infinity remains non-negative for any $\lambda$, the angular momentum flux changes sign at $\lambda=0$. This then also implies the sign change of the growth rate $\omega_I$. In Fig.~\ref{fig:testfieldcomp}, we have already observed this behavior. A natural result of this argument is that the unstable modes in the eikonal limit reduce precisely to zero-modes in the critical spacetime that contains an evanescent ergosurface \cite{Keir:2018hnv}. Away from the eikonal limit, only the co-rotating opposite-sign angular momentum modes are unstable. For systems with additional dissipation channel, the above energy-to-angular momentum relation of the unstable test field may no longer hold, and the zero-frequency mode may not signal the onset of the instability, as found, e.g., in Ref.~\cite{Maggio:2017ivp}.

\subsection{Overtones and angular modes}

\begin{figure}[t]
\includegraphics[width=0.48\textwidth]{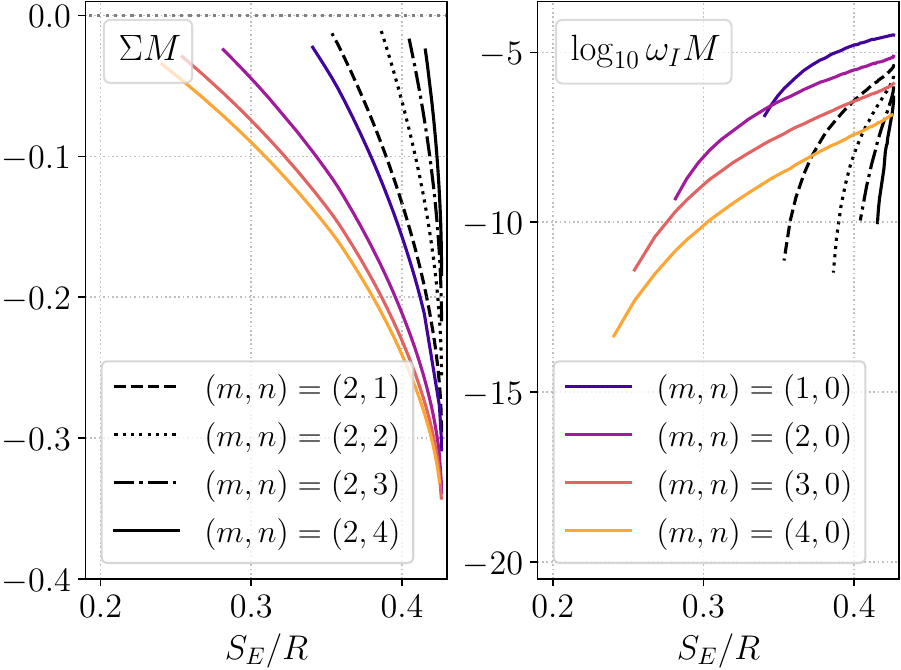}
\caption{The frequencies $\Sigma=\omega_R/m$ and growth rates $\omega_I$ of various ergoregion unstable $\ell=m$ modes along the family of $\tilde{m}=2$ KKLS star solutions parametrized by it's radial (coordinate) ergoregion size in the equatorial plane $S_E$. Recall, $R$ are the star's radii. We restrict to the branch of star solutions below the maximum mass.}
\label{fig:DIexample}
\end{figure}

Friedman showed, for any horizonless spacetime with compact ergoregion, there exists a $m_0$, such that all massless scalar modes with $m\geq m_0$ are unstable \cite{Friedman:1978}. This was demonstrated explicitly assuming $\ell=m$ and $n=0$. Conversely, along a sequence of spacetimes containing an ergoregion of increasing size, the critical azimuthal index $m_0$ decreases down towards unity. This behavior can already be seen in Figs.~\ref{fig:testfieldcomp} and is more apparent in Fig.~\ref{fig:DIexample}, for the lowest few $\ell=m$ modes along a sequence of BSs. Intuitively, for a given ergoregion, the large-$m$ modes are highly localized on the stable light ring inside the ergosphere, resulting in minimal overlap with the ergoregion's exterior, i.e., minimal ``leakage''. Hence, these eikonal modes radiate the least, which results in small growth rates. The low-$m$ modes, on the other hand, are associated with wavefunctions of spatial scales comparable to the size of the trapping potential. These have largest support in the ergoregion's exterior, radiate the most, and turn unstable only if the ergoregion's size is comparable to that of the trapping potential. 

Our results suggest that in addition to $m_0$ the set of unstable modes is also bounded by critical overtone and polar mode numbers: for each azimuthal index $m\geq m_0$ there exist a \textit{finite} number of unstable overtones $0\leq n\leq n_0$ and polar modes $\ell_0\geq\ell\geq m_0$. In Fig.~\ref{fig:DIexample} this behavior is shown for brevity only for the overtones and a particular sequence of bosonic stars. We find this to hold in all considered bosonic stars and polar modes. From there it is clear that for small ergoregions only the fundamental mode, $n=0$, is unstable, while more and more overtones, at fixed azimuthal index, turn unstable as the ergoregion grows in size. Note, the growth rates of higher overtones and polar modes are orders of magnitude smaller unless the background spacetime is deep in the large-ergoregion regime. Physically, larger overtones and polar mode numbers imply a delocalization of the wavefunction away from the stable light ring in the radial and polar directions, respectively. This increases the support of these states outside of (and minimizes the overlap with) the ergoregion, unless the latter is large compared to the size of the trapping potential. In particular, if the state is too delocalized from the ergoregion, i.e., $\ell>\ell_0$ or $n>n_0$, then its energy is positive, and hence, it decays in time. Note, a similar structure exists also for Kerr-like objects, see e.g., Ref.~\cite{Zhong:2022jke}.

\subsection{Small-frequency limit}

\begin{figure}
\includegraphics[width=0.48\textwidth]{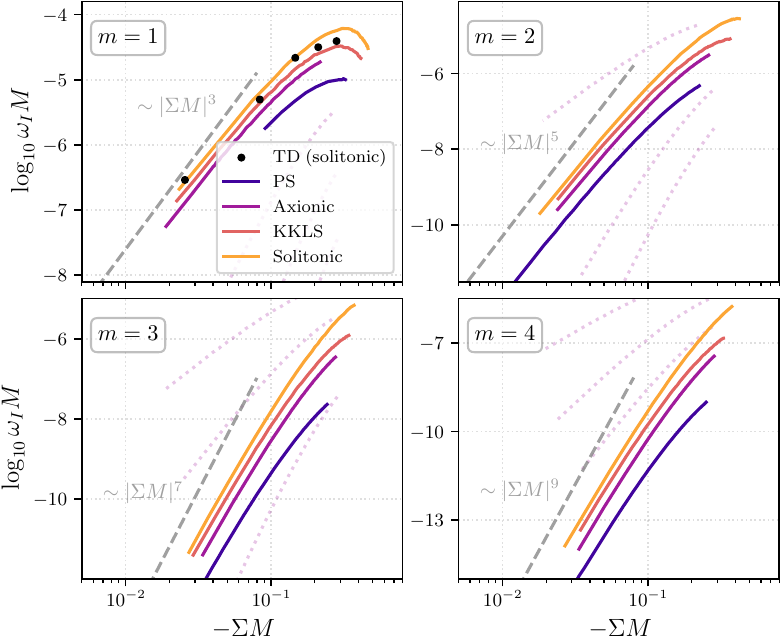}
\caption{The lowest four $m=\ell$ and $n=0$ scalar ergoregion unstable mode frequencies and growth rates along each of the four families of bosonic stars considered in this work (including solutions beyond the maximum mass of the associated family). Each panel corresponds to a different azimuthal index, while the dotted lines indicate the other azimuthal modes of the axionic family in each panel for reference. The gray dashed lines show the scaling behavior of the growth rates towards small frequencies as expected based on eq.~\eqref{eq:smallfrequencyexpansion}. For comparison, we also added the time-domain estimates [labeled ``TD (solitonic)''] shown in Fig.~\ref{fig:testfieldcomp}; the error bars were removed for visual clarity.}
\label{fig:allmodes}
\end{figure}

\begin{figure}
\includegraphics[width=0.48\textwidth]{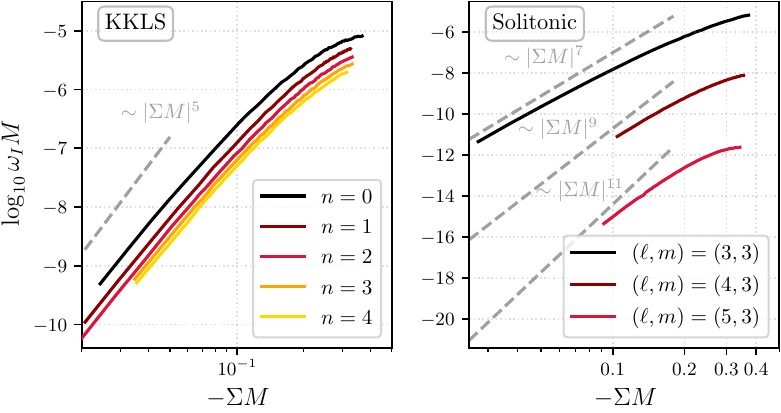}
\caption{(left) The fundamental $m=\ell=2$ unstable mode along the $\tilde{m}=2$ KKLS family of BSs together with the first four overtones. (right) Higher-order polar modes of the $m=3$ fundamental unstable modes along the $\tilde{m}=3$ family of solitonic stars (right panel); here $n=0$ in all three cases.}
\label{fig:overtones}
\end{figure}

The small-frequency limit of unstable modes along a sequence of all bosonic star spacetimes we find to exhibit explicit universal properties. In the case of Kerr-like objects with perfect reflectivity, it has been shown that in this regime, i.e., $-\Sigma M\ll 1$, the growth rates of $n=0$ modes follow the scaling \cite{Vilenkin:1978uc,Barausse:2018vdb,Maggio:2018ivz}
\begin{align}
\omega_I M\sim |\omega_RM|^{2\ell+1}.
\label{eq:smallfrequencyexpansion}
\end{align}
Here we show that this scaling applies also in all families of bosonic stars considered in this work. In Fig.~\ref{fig:allmodes}, we show the lowest four $\ell=m$ and $n=0$ massless scalar ergoregion unstable modes along all four families of star solutions. The growth rates approach the scaling \eqref{eq:smallfrequencyexpansion} towards the $-\Sigma M\ll 1$ limit. We also find overtones not to impact this leading scaling, as shown in the left panel of Fig.~\ref{fig:overtones}. In this case, the radial node number $n$ modifies the overall magnitude of the growth rates, while retaining \eqref{eq:smallfrequencyexpansion} towards small frequencies; explicitly, we find $\omega_I \approx (n+1)^{-1}\omega_I^{n=0}$. Recall, however, only a finite set of unstable overtones exists for any $\ell=m$. Lastly, in Fig.~\ref{fig:overtones} we also show the first two higher-order polar modes $\ell >m$ along a particular sequence of stars. We were unable to compute these modes down to small $-\Sigma M\lesssim 0.1$, therefore, while less conclusive, we find the scaling \eqref{eq:smallfrequencyexpansion} to apply roughly also to these. 

Towards large frequencies, $-\Sigma M\gtrsim 0.1$, higher-order corrections to \eqref{eq:smallfrequencyexpansion} become important. The larger the ergoregion, the stronger the associated frame-dragging effects become, which implies an increase in the real frequency $-\Sigma$ of the unstable field configurations. Following the CFS argument, the mode remains unstable so long as it counter-rotates in the co-rotating frame of the object sourcing the spacetime. If the ergoregion, and hence, the frequency $-\Sigma$, increase sufficiently, then the formerly unstable mode begins to co-rotate in the co-rotating frame of the object, and consequently, turns stable. The turnover of the $m=1$ modes shown in Fig.~\ref{fig:allmodes} around $-\Sigma M\approx 0.3$ is likely due to the modes approaching this co-rotation point. While for bosonic stars no unique definition of an orbital period exists (see e.g., Refs.~\cite{Ryan:1996nk,Siemonsen:2020hcg,Adam:2022nlq}), for Kerr-like objects the horizon frequency $\Omega_H$ is a natural candidate; indeed, in this case, the growth rates exhibit precisely such a turnover \cite{Vilenkin:1978uc}: $\omega_I\propto \Sigma-\Omega_H$.\footnote{Recall, in our convention $\Sigma,\Omega_H<0$.}

\subsection{Eikonal limit}
\label{sec:eikonal}

Comins and Schutz showed in Ref.~\cite{Comins:1978}, the growth timescale of the instability increases exponentially with azimuthal mode number $m=\ell$ and $n=0$ towards the eikonal limit: $\omega_I \sim e^{-\beta m}$. There it was furthermore suggested that the (appropriately normalized) frequency of the unstable modes should approach the orbital frequency of null geodesics residing at the equatorial stable light ring. The radial potential for equatorial null geodesics of energy $E$ and angular momentum $L$ is precisely of the form \eqref{eq:eikonalpot} with $\omega_R\rightarrow E$ and $m\rightarrow L$. The impact parameter $b^{-1}=E/L=\omega_-$ of the counter-rotating equatorial light ring is then related to the frequency of the ergoregion unstable massless frequencies
\begin{align}
\omega_R\approx \ell \omega_-,
\label{eq:eikonallimit}
\end{align}
in the $\ell=m\gg 1$ limit. Here we demonstrate this relationship explicitly. In Fig.~\ref{fig:testfieldcomp}, we compare the unstable mode frequencies $\Sigma$ for the first few families of $\ell=m$ modes against the counter-rotating null geodesic frequency $\omega_-$ (there simply labeled ``Eikonal limit''). With increasing $\ell=m$, the frequencies $\Sigma$ approach $\omega_-$, confirming relation \eqref{eq:eikonallimit}. Naturally, as these null geodesics are stably trapped, the imaginary part vanishes in the eikonal limit, $\omega_I\rightarrow 0$. An analogous relation exists in non-spinning spacetimes with stable light ring. There the frequency of stably quasi-trapped \textit{decaying} massless modes approaches the orbital frequency of trapped null geodesics, see e.g., Ref.~\cite{Cardoso:2014sna}. Similarly, in this setting unstably trapped null geodesics are linked to both the real and imaginary parts of decaying quasi-normal modes \cite{Goe1972,Ferrari:1984zz}.

\subsection{Comparison to Kerr-like objects}

\begin{figure}
\includegraphics[width=0.99\linewidth]{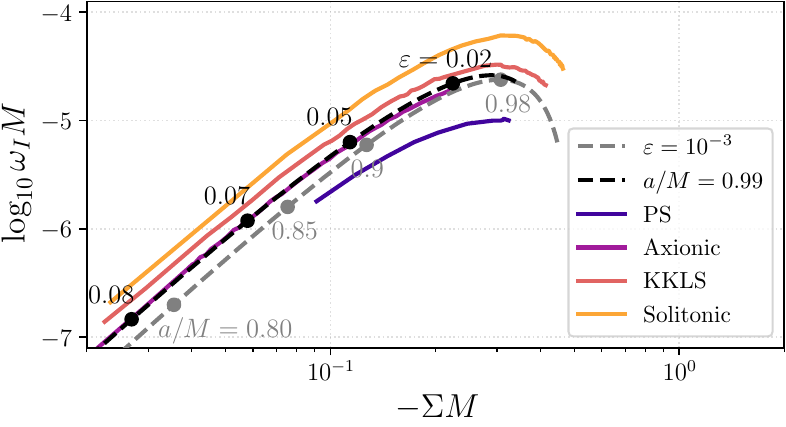}
\caption{The zeroth overtone of the $\ell=m=1$ ergoregion unstable modes along the various families of bosonic stars (solid lines) with the same labels as in Fig.~\ref{fig:allmodes}. These are compared with the corresponding frequencies and growth rates of the equivalent unstable modes in a Kerr-like object of spin $a/M=0.99$ for a series of $\varepsilon\geq 10^{-3}$ (dashed black line), and with $\varepsilon=10^{-3}$ for a series of spins $a/M<1$ (dashed gray line); these were obtained in Ref.~\cite{Zhong:2022jke} and Ref.~\cite{Maggio:2017ivp}, respectively. As in Fig.~\ref{fig:allmodes}, this also includes bosonic stars beyond the maximum mass of their families of solutions.}
\label{fig:kerr_comp}
\end{figure}

Linear waves propagating on the Kerr exterior of spin parameter $a$ and mass $M$ with boundary conditions imposed at the Boyer-Lindquist radius $r_{\rm obj}=r_+(1+\varepsilon)$, with $r_+=M+(M^2-a^2)^{1/2}$ and $\varepsilon>0$, is referred to as a Kerr-like object. In particular, a massless scalar test field with perfectly reflecting boundary conditions at $r_{\rm obj}$ exhibits an ergoregion instability if $|a|\neq 0$ and $r_{\rm obj}<2M$ (see e.g., Ref.~\cite{Zhong:2022jke}). The parameter $\varepsilon$ controls the size of the ergoregion (at fixed spin $a$) analogous to $\mu M$ and $S_E/R$ in Figs.~\ref{fig:testfieldcomp} and~\ref{fig:DIexample}, respectively, parametrizing each family of bosonic star solutions.

In Fig.~\ref{fig:kerr_comp}, we compare the unstable mode frequencies along each considered family of bosonic stars to those of a massless scalar test field around a Kerr-like object at fixed spin $a/M=0.99$, but varying $\varepsilon$. In addition to the small-frequency scaling, even the absolute value of the growth rates (for a given frequency) agree to within a factor of two, which is also comparable to the systematic uncertainty of the direct integration method (see Figs.~\ref{fig:testfieldcomp} and~\ref{fig:allmodes}). This agreement continuous to hold also towards large frequencies, and even the turnover around $-\Sigma M\approx 0.3$ agrees remarkably well. Recall from Fig.~\ref{fig:family}, the spin of bosonic stars changes along the families of solutions, but is roughly around $J\approx M^2$ in the range relevant for the comparison in Fig.~\ref{fig:kerr_comp}. This make the comparison to a near-extremal Kerr-like object most appropriate. Nonetheless, since the spin is fixed in the latter, but not in the former, the two are not strictly comparable.\footnote{The spin of Kerr-like objects has an impact on this relation in the small-frequency limit~\cite{Barausse:2018vdb,Maggio:2018ivz}.} 

This agreement seems surprising, as bosonic stars are a priori fundamentally different from the Kerr spacetime. However, it may be understood as follows: in Ref.~\cite{Siemonsen:2024snb}, it was observed that the frequency of the unstable counter-rotating light ring in bosonic stars approaches the near-extremal Kerr value in the high-compactness limit of the former. It has been explicitly shown that sequences of matter distributions exist, which source the (extremal) Kerr spacetime in a high-compactness limit \cite{Israel:1970kp,Bar1971,Ansorg:2002vh,Fischer:2005uy} (see also Refs.~\cite{Mazza:2021rgq,Bianchi:2024shc}), and that multipole moments of horizonless compact objects approach those of spinning black holes in the high-compactness limit~\cite{Yagi:2015upa,Pani:2015tga,Raposo:2018xkf,Glampedakis:2017cgd}. This leads us to conjecture that such a Kerr-limit exists also for bosonic stars (potentially, in the $\tilde{m}\rightarrow \infty$ and high-compactness limits). This would naturally explain the observations made in Ref.~\cite{Siemonsen:2024snb} and the agreement found in Fig.~\ref{fig:kerr_comp}, particularly that $\Omega_H$ dictates the turnover around $-\Sigma M\approx 0.3$. 

\section{Weakly nonlinear effects}
\label{sec:nonlinear}

Thus far, we have focused on the ergoregion instability of a massless scalar test field. In an effort to go beyond the linear level, and begin understanding the impact of backreaction of the instability on the spacetime, here we discuss leading nonlinearities. These effects allow for speculation about the late-time nonlinear development of the instability. In particular, a slow-down of the instability at the weakly nonlinear level suggests that the saturation of the unstable process is self-regulating, whereas an enhancement of the instability implies that strongly nonlinear effects will dictate the final fate of the system, potentially leading to a drastic departure from the original stationary spacetime. In the following, we begin by identifying a weakly nonlinear shift in the unstable modes' complex frequencies and additional radiation channels (such as gravitational waves) as primary leading effects impacting the evolution of the instability. As an example, we proceed by quantifying these effects in a $\tilde{m}=1$ PS solution. In this case, we find evidence that leading nonlinearities amplify, rather than weaken, the unstable process, suggesting that strongly nonlinear effects ultimately decide the fate of the instability.

\subsection{Leading nonlinearities} \label{sec:leadingnonlin}

To understand the backreaction, we expand all relevant field equations to second order in the amplitude $\varepsilon$ of the perturbing scalar probe field $\Phi\sim \varepsilon$. To that end, the focus lies on PS backgrounds, though, the following arguments carry over also to BS backgrounds. At linear order in $\varepsilon$, the probe field acts as a test field and the behavior is as described above. At second order, it results in a non-vanishing stress-energy tensor [see eq.~\eqref{eq:setensor}]
\begin{align}
T_{\mu\nu}\sim \varepsilon^2.
\end{align}
This sources the linearized coupled Einstein-Proca equations describing perturbations of a PS background. Let $A_\mu^{\rm PS}$ and $g^{\rm PS}_{\mu\nu}$ be the vector field and metric solutions describing a given PS background, and $a_\mu$ as well as $h_{\mu\nu}$ be their leading perturbations, respectively; then the linearized Einstein-Proca equations read
\begin{align}
\begin{aligned}
G^{(1)}_{\alpha\beta}[h]=8\pi (T_{\alpha\beta}+T_{\alpha\beta}^{(1)}[h,a]), \\
\nabla_\alpha f^{\alpha\beta}+\frac{1}{2}F^{\alpha\beta}_{\rm PS}\nabla_\alpha h=\mu^2 a^\beta,
\end{aligned}
\label{eq:linearizedeq}
\end{align}
where $f^{\mu\nu}$ is the field strength associated with $a_\mu$ and $G^{(1)}_{\alpha\beta}[h]$ as well as $T^{(1)}_{\alpha\beta}[h,a]$ are the leading perturbation of the Einstein and background stress energy tensors~\cite{Misner:1973prb}, respectively. The backreaction of $h_{\mu\nu}, a_\mu$ onto $\Phi$ occurs at third order in $\varepsilon$, i.e., the probe field's equations are $\square_{g^{\rm PS}} \Phi=\mathcal{O}(\varepsilon^3)$. In this linearized context, there likely also exists a gravitational wave-driven ergoregion instability akin to $w$-modes in fluid stars; we ignore this here, and focus entirely on the backreaction of the scalar ergoregion instability.

This implies that at order $\varepsilon^2$ the probe field $\Phi$ continues to grow exponentially, and only $h_{\mu\nu}$ and $a_\mu$ tap into the energy and angular momentum of the background spacetime, $\dot{E}_\Phi=2\omega_I E_\Phi$,
while the latter implies
\begin{align}
\dot{E}_{h+a}= & \ T_\Phi-P_{\rm GW}-\dot{E}_{a}.
\label{eq:linenergyevo}
\end{align}
Here $\dot{E}_{h+a}$ is the energy contained in the leading perturbations of the background PS, $T_\Phi$ and is the transfer rates of energy from the probe field $\Phi$ to these leading perturbations, $P_{\rm GW}$ is the emitted gravitational wave power, and $\dot{E}_{ a}$ the total outgoing massive vector energy flux at infinity. Recall from above that $P_{\rm GW},\dot{E}_{a}\sim h^2,a^2\sim\varepsilon^4\sim E_\Phi^2/M^2$. The test field ergoregion instability is driven by positive energy fluxes at future null infinity; hence, any additional dissipation mechanisms tapping into the negative energy of $\Phi$---such as $P_{\rm GW},\dot{E}_{a}> 0$---will amplify the instability, rather than weaken it.

The unstable process of the probe field $\Phi$ is affected only at the $\sim\varepsilon^3$ level. Leading metric and vector perturbations $h_{\mu\nu}$ and $a_\mu$ result in effective scalar self-interactions, which modify the unstable modes' frequency and growth rate away from their test field values. As a result, this effect scales as $h,a\sim \varepsilon^2\sim E_\Phi/M\sim J_\Phi/J$ in terms of the test field's energy and angular momentum $E_\Phi$ and $J_\Phi$, respectively. This is analogous to findings reported in Ref.~\cite{Siemonsen:2025fne} in a nonlinear scalar theory, as well as to weakly nonlinear gravitational effects around black holes~\cite{Baryakhtar:2017ngi,May:2024npn,Sberna:2021eui}.

\subsection{Example: Proca star} \label{sec:example_ps}

\begin{figure}[t]
\includegraphics[width=0.48\textwidth]{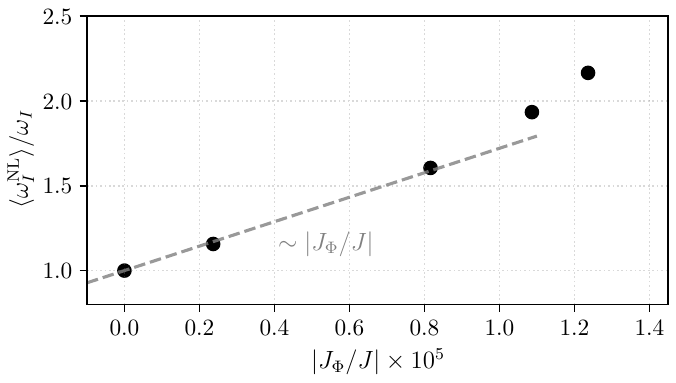}
\caption{Weakly nonlinear enhancement of the growth rate, $\omega^{\rm NL}_I=-\dot{J}_\Phi/(2J_\Phi)$, as a function of the amplitude of the perturbation (measured by its angular momentum $J_\Phi$), compared with its test field value $\omega_I$. The dashed line is a fit through the two data points associated with the smallest angular momentum $J_\Phi$ and fixed at the test field value. Brackets $\langle\dots\rangle$ indicate a rolling time-average as discussed in more detail in Appendix~\ref{app:gwpower}.}
\label{fig:ps_wnl}
\end{figure}

Having identified the two primary weakly nonlinear effects, i.e., a shift of the unstable mode' complex frequency and additional radiation channels, we focus on a single representative of the $\tilde{m}=1$ family of PSs with ergoregion to explicitly quantify these effects. To that end, we turn to fully nonlinear numerical evolutions of the combined Einstein-Proca-Klein-Gordon system governing the dynamics of the PS perturbed by the scalar field. The PS of interest has mass $\mu M=1.1238$ and frequency $\tilde{\omega}/\mu=0.555$. On this background spacetime, the most unstable scalar configuration has azimuthal index $m=2$ with frequency $-\Sigma M=0.028$ and growth rate $\omega_IM=1.3\times 10^{-10}$ (obtained using the time-domain method). To measure the impact of weakly nonlinear effects, we initialize the scalar field in this most unstable field configuration as initial data, but vary its initial amplitude. For each amplitude we measure the instantaneous growth rate using the outgoing angular momentum flux $\omega^{\rm NL}_I=-\dot{J}_\Phi/(2J_\Phi)$. These evolutions are performed enforcing axisymmetry on the metric and fixing the azimuthal index $m$ of the complex scalar field. To determine the amplitude of the gravitational wave emission, we map the complex scalar field $\Phi$ to a real field $\phi$, i.e., $\phi=\sqrt{2}\text{Re}(\Phi)$, and initialize $\phi$ at different amplitudes on the PS. We then proceed to perform these evolutions in three spatial dimensions to capture the non-axisymmetric nature of the emitted gravitational waves. Importantly, the initialization of the scalar field at weakly nonlinear amplitudes may cause spurious mode excitations within the star compared with the field configuration that grew to this amplitude through the unstable process. Thus, this approach should be interpreted as providing evidence for the broad behavior of the weakly nonlinear effects, rather than accurate predictions. In Appendix~\ref{app:NumSetupNon}, we provide further details on the numerical methods and convergence. 

In Fig.~\ref{fig:ps_wnl}, we show the instantaneous growth rate $\langle\omega^{\rm NL}_I\rangle$, compared to its test field counterpart $\omega_I$, as a function of angular momentum contained in the field configuration. As evident from there, weakly nonlinear gravitational effects serve to enhance the growth rate, rather than weaken the unstable process. Secondly, given the accuracy of our methods, we are able only to place an upper bound on the emitted gravitational wave power $P_{\rm GW}$. To that end, we define the angular momentum available for extraction by the $m=2$ scalar field configuration as the difference between the angular momentum of the above considered PS, $J$, and angular momentum of the PS solution residing at the zero mode of the family of $m=2$ scalar field configurations, $J_c$. This critical solution is found by extrapolating the linear frequency of the unstable modes $\Sigma$ along the family of PS solutions to $\Sigma=0$ (see Appendix~\ref{app:gwpower} for details). We assume that strongly nonlinear effects become important, as soon as the unstable field configuration contains at least half of the available angular momentum, $J_\Phi^{\rm strong}=(J-J_c)/2$.\footnote{The angular momentum $J^{\rm strong}_\Phi/J\approx 2.6\times 10^{-4}$ explains the smallness of the ratios $|J_\Phi/J|$ in Fig.~\ref{fig:ps_wnl}.} Scaling both our upper bound on $P_{\rm GW}$ and $\dot{E}_\Phi$ to the amplitude $J_\Phi^{\rm strong}$, we find that gravitational radiation is suppressed before strongly nonlinear effects become important:
\begin{align}
P_{\rm GW}|_{\rm strong}\lesssim \dot{E}_\Phi|_{\rm strong}.
\label{eq:gwupperbound}
\end{align}
Lastly, the linearized vector wave equation \eqref{eq:linearizedeq} is sourced by perturbations with frequency $2\omega_R\pm \tilde{\omega}$, which leads to asymptotically free states (i.e., a radiating Proca field) only if this frequency surpasses the vector mass $\mu$. Combining the $\tilde{\omega}$ from above and $\omega_R=2\Sigma$, we find $|2\omega_R\pm\tilde{\omega}| < \mu$, which implies $\dot{E}_a=0$. 

\section{Conclusion}

In this work, we systematically investigated the ergoregion instability of a massless scalar test field on a large variety of bosonic star background spacetimes using three different methods (a WKB, a frequency-domain, and a time-domain method). Broad features of the scalar mode structure are captured by all three techniques, while in specific cases, predictions from the WKB and frequency-domain methods (based on the slow-rotation approximation) can be orders of magnitude away from the true values. Nonetheless, we found the frequency-domain method is reasonably accurate particularly away from the small-frequency limit. 

Overall, we found that all modes enter the unstable regime through a zero-mode and briefly review its physical origin. Our results suggest that there exist only a finite number of unstable overtones, $n_0\geq n\geq 0$, and polar modes, $\ell_0\geq\ell\geq m$, for a given unstable $m=\ell$ and $n=0$ field configuration. In the small-frequency limit the growth rates universally scale as $\omega_IM\sim |\omega_R M|^{2\ell+1}$ as found for Kerr-like objects~\cite{Vilenkin:1978uc,Barausse:2018vdb,Maggio:2018ivz}. We found this scaling to hold also for unstable overtones and high-order polar modes $\ell>m$. Corrections from the scaling become important towards high frequencies and eventually lead to a turnover of the growth rate as each mode becomes stable in the high-frequency regime. In the eikonal limit, we explicitly demonstrated that the frequency of unstable $\ell=m$ modes approaches the orbital frequency $\omega_-$ of null geodesics residing in the counter-rotating stable light ring: $\omega_R\rightarrow \ell \omega_-$. Interestingly, we uncovered a near-universal behavior of the growth rates and frequencies across all considered bosonic stars also away from the small-frequency limit. A direct comparison of the unstable states on bosonic star backgrounds to frequencies and growth rates of massless scalar unstable modes in Kerr-like objects revealed a remarkable agreement. From small to high frequencies, the difference between the former and latter is roughly within the uncertainties of our methods. This suggests the existence of a Kerr spacetime limit of bosonic stars, at least in the near-extremal range, potentially explaining this near-universal behavior. This also implies that the fastest e-folding time for the scalar ergoregion instability in bosonic stars is $\tau\gtrsim 10^4 M$. 

Furthermore, we considered the leading backreaction of the unstable massless scalar field on the background bosonic stars. Generally, the resulting linear perturbations of the star can source additional emission channels at infinity and may induce a weakly nonlinear shift of the linear frequencies and growth rates. In the context of a highly compact spinning Proca star we found indications that this nonlinear shift in rates leads to an enhancement of the unstable process, while additional radiation channels are insignificant for the growth phase of the instability. This suggests that strongly nonlinear gravitational effects dictate the final fate of the instability in full General Relativity. 

Finally, in our three-dimensional numerical relativity evolutions of ultracompact spinning Proca stars with stable light rings, we find no evidence for an unstable process as conjectured to exist in Refs.~\cite{Keir:2014oka,Cardoso:2014sna}. In our setting, these star solutions are perturbed by ergoregion unstable scalar field configurations with large support in the stable light ring. The amplitude of this perturbing field is chosen to be marginally below levels, where spurious numerical instabilities are induced or immediate black hole formation occurs. Despite this, we are unable to reproduce the findings of Ref.~\cite{Cunha:2022gde}, albeit our evolutions are at best comparable in length to instability timescales quoted there. This is consistent with observations reported in Ref.~\cite{Evstafyeva:2025mvx} (see also Refs.~\cite{Siemonsen:2024snb,Marks:2025jpt}), arriving at similar conclusions.

A future direction is to explore the gravitationally-driven linear ergoregion instability in these bosonic star solutions, akin to $w$-modes in fluid stars, as well as consider other rapidly spinning horizonless compact objects~\cite{Danielsson:2023onu,Giri:2024cks} in an effort to study the nonlinear gravitational saturation of this instability in light of Ref.~\cite{Siemonsen:2025fne,Siemonsen:2025ucx}. 

\begin{acknowledgments}
We are grateful for many valuable discussions with Will East throughout the development of this work. We also thank Elisa Maggio for interesting exchanges and Zhen Zhong for sharing the data of Ref.~\cite{Zhong:2022jke}. This work used \texttt{anvil} at Purdue University through allocation PHY250024 from the Advanced Cyberinfrastructure Coordination Ecosystem: Services \& Support (ACCESS) program \cite{access}, which is supported by National Science Foundation (NSF) Grants OAC-2138259, -2138286, -2138307, -2137603, and -2138296. This research was enabled in part by support provided by the Digital Research Alliance of Canada (alliancecan.ca). Calculations were performed on the Symmetry cluster at Perimeter Institute and the Narval cluster at the École de technologie supérieure in Montreal.
\end{acknowledgments}

\appendix
\section{Numerical evolution scheme} \label{app:NumSetup}

\subsection{Test field evolution} \label{app:NumSetupTest}

Here we briefly outline the test field time-domain methods used throughout this work; That is, we begin with the evolution scheme used to solve the Klein-Gordon equation \eqref{eq:KGequation} on the spinning bosonic star backgrounds making no approximations, as discussed in sec.~\ref{sec:td_method1}. To that end, we make use of the framework introduced in Refs.~\cite{Siemonsen:2020hcg,Pretorius:2004jg}. In particular, the scalar field equations are discretized using fourth-order accurate finite differences together with fourth-order accurate Runge-Kutta integration in time. We use fixed mesh refinement centered on the star with refinement ratio $2:1$. Crucially, we assume the scalar field's azimuthal dependence follows $\partial_\varphi\Phi=im\Phi$, i.e., we make use of the generalized Cartoon Method (see e.g., Ref.~\cite{Pretorius:2004jg}). To initialize the scalar field as closely as possible to the most unstable field configuration, we use the $m=1$ frequency-domain solutions (discussed in sec.~\ref{sec:di_method}) for a given background. 

\begin{figure}[t]
\includegraphics[width=1\linewidth]{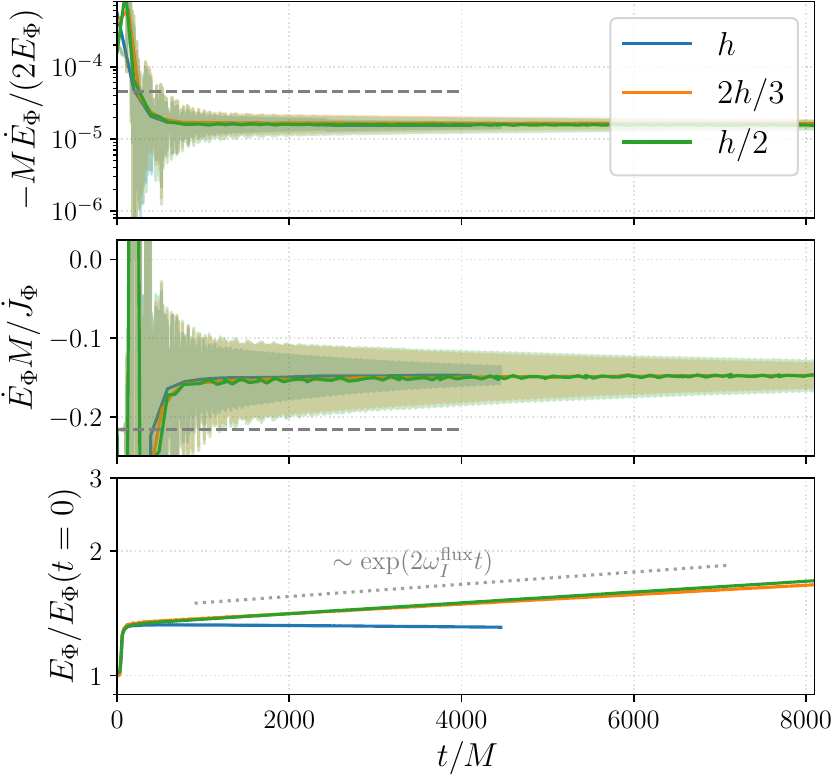}
\caption{The shaded lines show the evolution of the real and imaginary parts of the frequency, as well as the scalar energy $E_\Phi$, of the ergoregion unstable $m=1$ massless scalar mode on the $\tilde{\omega}/\mu\approx 0.427$ and $\tilde{m}=3$ solitonic BS spacetime, depending on the grid spacing $h$ used; solid lines are rolling time-averages. Dashed gray lines indicate the real and imaginary parts of the $m=1$ unstable mode's frequency obtained from the direct integration method. Lastly, we compare $\omega_I^{\rm flux}=-\dot{E}_\Phi/(2E_\Phi)$ with the exponential growth of $E_\Phi$ in the bottom panel.}
\label{fig:scalar_conv}
\end{figure}

As an example of the dynamics, in Fig.~\ref{fig:scalar_conv}, we show on the evolution of the massless scalar field on the $\tilde{\omega}/\mu\approx 0.427$ and $\tilde{m}=3$ solitonic boson star spacetime across three resolutions (i.e., the central data point in Fig.~\ref{fig:testfieldcomp}). In particular, we present the evolution of the energy functional $E_\Phi$, defined in \eqref{eq:energyfunctional} (with $S\subset\Sigma_t$ the coordinate ball of radius $100M$), and the real and imaginary parts of the linear frequency, $\omega_R M\approx \dot{E}_\Phi M/\dot{J}_\Phi$ and $\omega_IM\approx -M\dot{E}_\Phi/(2E_\Phi)$, respectively. An initial transient of the scalar field leads to a short burst of scalar radiation. Therefore, at early times, $t\lesssim 100M$, both the energy $E_\Phi$ and scalar fluxes grow exponentially in time with roughly the rate predicted by the direct integration method. Once the burst of scalar radiation left the domain $S$, the energy settles into a new growing state with a different growth rate; this can be seen in the up tick of $-E_\Phi$ in the bottom panel of Fig.~\ref{fig:scalar_conv} around $t/M\approx 100$. Specifically, the fluxes -- and frequency and growth rate -- begin to assume a new value after a few light crossing times, $\sim\mathcal{O}(10^2)M$. Due to computational cost, we evolve the system only for a fraction of the scalar field's e-folding time, $\omega_I^{-1}/M\approx 5\times 10^4$. We checked explicitly that this field configuration, restricted to the coordinate sphere intersecting the ergoregion roughly at the center in the equatorial plane, has largest support over the $\ell=m=1$ \textit{spherical} harmonic (with exponentially smaller support over the $\ell>m$ spherical harmonics in these coordinates). We estimate the error on $\omega_R$ by measuring the amplitude of the oscillations of $\dot{E}_\Phi M/\dot{J}_\Phi$ around the central value. The error estimate for $\omega_I$ is obtained from the difference between $-M\dot{E}_\Phi/(2E_\Phi)$ and an exponential fit to $-E_\Phi$. This difference is comparable to, or larger, than the amplitude of the oscillations of $-M\dot{E}_\Phi/(2E_\Phi)$ (shown as the shaded regions in Fig.~\ref{fig:scalar_conv}). All data points shown in Fig.~\ref{fig:testfieldcomp} were obtained from evolutions with default grid spacing of $h/M\approx 9.6\times 10^{-3}$ on the finest level.

\subsection{Fully nonlinear evolution} \label{app:NumSetupNon}

The fully nonlinear evolution of the Einstein-Proca-Klein-Gordon system of equations governing the perturbed PSs presented in sec.~\ref{sec:example_ps} were performed using framework developed in Refs.~\cite{Pretorius:2004jg,Siemonsen:2020hcg}, but for reasons discussed in \cite{Evstafyeva:2025mvx} employs the Z4 formulation~\cite{Bona:2003fj}, with more details provided in \cite{Siemonsen:2025ucx}; that is, we find gauge instabilities to emerge for highly relativistic such star solutions within the generalized harmonic formulation of the Einstein equations and stationary gauge, in direct analogy to what was found in Ref.~\cite{Evstafyeva:2025mvx}. Therefore, all evolutions are performed in the standard moving puncture gauge~\cite{Bona:1994dr,Baker:2005vv,Campanelli:2005dd,Alcubierre:2001vm}. As before, we employ fourth order accurate finite difference stencils to discretize the equations in conjunction with Runge-Kutta temporal integration and a grid with fixed mesh refinement with ten levels centered on the star. The metric and Proca initial data used here is constructed numerically using methods from Ref.~\cite{Siemonsen:2023hko}. The massless scalar field serves as the ergoregion unstable probe field perturbing the star solution. The complex scalar field is first evolved enforcing the axisymmetry of the background spacetime starting from initial data obtained using the direct integration method (see sec.~\ref{sec:di_method}). After the latter settled into a new quasi-stationary field configuration, the final state serves as initial data to measure the weakly nonlinear frequency shift and gravitational wave emissions. For the nonlinear shift in the linear instability growth rates, we utilize this dimensionally reduced setting, while for the extraction of the weakly nonlinear gravitational wave power, these scalar field data are mapped into initial data in three dimensions using $\partial_\varphi\Phi=2i\Phi$ and defining a purely real scalar probe field, $\phi= \sqrt{2}\text{Re}(\Phi)$. 

\begin{figure}[t]
\includegraphics[width=1\linewidth]{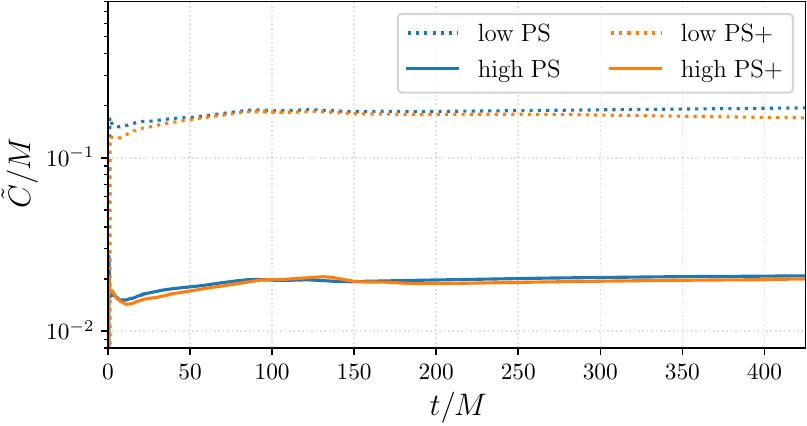}
\caption{Evolution of the integrated violations, $\tilde{C}=\int_S d^3x \sqrt{\gamma}(\mathcal{H}^2+\sum_{i=1}^3\mathcal{M}^2_i)^{1/2}$, of the Hamiltonian, $\mathcal{H}$, and momentum constraint violations, $\mathcal{M}_i$. The low and high resolutions correspond to a grid spacing of $h/M\approx 1.3\times 10^{-2}$ and $h/2$ on the finest level, respectively. We show both the evolution of the unperturbed PSs (labeled ``low PS'' and ``high PS'') and the perturbed stars (labeled ``low PS+'' and ``high PS+'').}
\label{fig:vector_cnst_conv}
\end{figure}

As a test of our methods, we track the convergence of the constraint violations in two different three-dimensional settings: (a) the evolution of the $\tilde{m}=1$ PS of mass $\mu M=1.1238$ and frequency $\tilde{\omega}/\mu=0.555$ without ergoregion unstable scalar probe field, and (b) the evolution of the same star with real scalar probe field $\phi$ with amplitude $|J_\phi/J| =2.9\times 10^{-5}$. The convergence of these constraint violations are shown in Fig.~\ref{fig:vector_cnst_conv}. In both cases, the violations converge to zero roughly at the expected fourth order; in case (b), the constraint violations introduced by the probe field are below those of the truncation error of the star's evolution as measured by $\tilde{C}$ defined in Fig.~\ref{fig:vector_cnst_conv}. Crucially, we observed spurious non-converging exponential growth of constraint violations or prompt collapse to a black hole for initial probe field amplitudes with $|J_\phi/J| \gtrsim 8\times 10^{-5}$. 

Lastly, we comment on the presence of a ``light ring instability'' as claimed to exist in Ref.~\cite{Cunha:2022gde}. There, a nonlinear drift of Proca star solutions with $\tilde{\omega}/\mu\geq 0.66$ was observed on timescales $t\mu\lesssim 10^3$ and attributed to a nonlinear mechanism active in the stable light ring. The Proca star solution analyzed here, is more compact, i.e., $\tilde{\omega}/\mu=0.555$, and hence, exhibits a larger stable light ring and an ergoregion; in this setting, one may naively expect shorter timescales associated with the nonlinear instability. Furthermore, while in Ref.~\cite{Cunha:2022gde} the instability was excited only with truncation error of their numerical scheme, here we perturb the star solution with the maximal allowed amplitude before spurious numerical instabilities appear or the solution promptly collapses to a black hole. Despite this, we observe no sign of a drift or an otherwise nonlinear process on the timescales shown in Fig.~\ref{fig:vector_cnst_conv} (recall $t=400M\approx 450/\mu$ for this star); consistent with findings reported in Ref.~\cite{Evstafyeva:2025mvx}. These timescales are roughly comparable, though likely slightly shorter, than the naive extrapolation of timescales quoted in Ref.~\cite{Cunha:2022gde} (see their Fig.~2) to the more compact stars studied here.

\section{Weakly nonlinear effects} \label{app:gwpower}

\begin{figure}[t]
\includegraphics[width=0.48\textwidth]{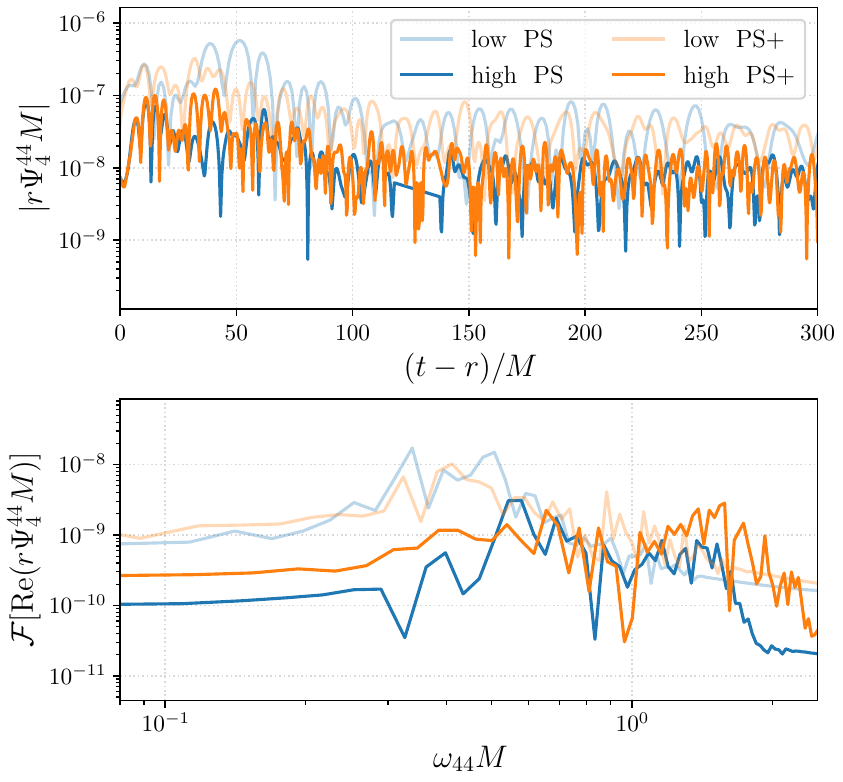}
\caption{The amplitude as a function of time (top) and the Fourier transform as a function of frequency (bottom) of the $(\ell,m)=(4,4)$ $s=-2$ spin-weighted spherical harmonic component of the Newman-Penrose scalar $\Psi_4$ extracted on a coordinate sphere with radius $r=100M$. The labels are the same as in Fig.~\ref{fig:vector_cnst_conv}. The Fourier transform is computed from the timeseries of $\Psi^{44}_4$ between $t-r=150M$ and $t-r=300M$. Notice, the ``high PS'' case is missing data roughly in the range $(t-r)/M\in [115,140]$ with no further significance for the analysis.}
\label{fig:gw_upper_bound}
\end{figure}

To determine the impact of weakly nonlinear effects on the instability's growth rate, we restrict to axisymmetric numerical evolutions, and consider the backreaction of the complex scalar field with azimuthal index $m=2$ as constructed in Appendix~\ref{app:NumSetupNon}. These test field initial data are then scaled to angular momenta shown in Fig.~\ref{fig:ps_wnl}. Given the PS initial data, we solve the system of Einstein-Proca-Klein-Gordon equations as detailed above and extract the growth rate from the radiation as $\omega^{\rm NL}_I=-\dot{J}_\Phi/(2J_\Phi)$. The backreaction of the scalar field on the star induces small oscillations of $\omega^{\rm NL}_I$ on timescales of $\approx 500 M$ (increasing in amplitude with increasing $J_\Phi$), particularly for $|J_\Phi/J|>0.8\times 10^{-5}$ (see Fig.~\ref{fig:ps_wnl}). This is averaged over as indicated by $\langle\dots\rangle$ in Fig.~\ref{fig:ps_wnl}. 

\begin{figure}[t]
\includegraphics[width=0.48\textwidth]{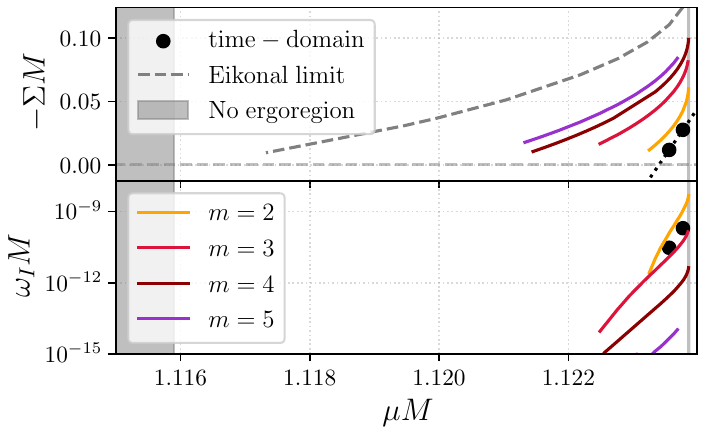}\hfill
\caption{The pattern speed $\Sigma=\omega_R/m$ and growth rate $\omega_I$ (obtained using the direct integration method) of the scalar ergoregion instability along the $\tilde{m}=1$ family of PSs. The $m=1$ mode becomes unstable only beyond the maximum mass of this family indicated by the vertical solid gray line. The time-domain estimates for the frequency and growth rate of the star introduced in the main text are labeled ``time-domain''. The black dotted line indicates the linear extrapolation of the two data points to towards $\Sigma=0$.}
\label{fig:testfieldcomp_ps}
\end{figure}

Turning to the upper bounds on the emitted gravitational wave power, in the previous appendix, we outlined some of the convergence properties of the (un)perturbed PS spacetime. Here we use these three-dimensional evolutions to place an upper bound on the total emitted gravitational wave energy flux, $P_{\rm GW}$. In particular, the $m=2$ ergoregion unstable real scalar field configuration, $\phi$, sources gravitational perturbations with $m_{\rm GW}=2m=4$ and frequency $\omega_{\rm GW}=|2\omega_R|=|4\Sigma|\approx 0.11/M$. Within our numerical setup, we compute the Newman-Penrose scalar $\Psi_4$, capturing the outgoing gravitational radiation. Therefore, and as long as the perturbing field $\phi$ is in the weakly nonlinear regime, the total flux reads
\begin{align}
P_{\rm GW}=\frac{|r\Psi_4^{44}|^2}{64\pi^2 \omega_R^2},
\end{align}
where we extract $\Psi_4$ on a coordinate sphere of radius $r=100M$, on which we project $\Psi_4$ using $s=-2$ spin-weighted spherical harmonics, to obtain the $(\ell,m)=(4,4)$ mode of the Newman-Penrose scalar: $\Psi_4^{44}$. For the PS considered in the main text (and discussed in Appendix~\ref{app:NumSetupNon}), the $m=2$ scalar probe field leads to a frequency $|\omega_{\rm GW}M|=0.11$. 

In Fig.~\ref{fig:gw_upper_bound}, we show both the amplitude and frequency of $\Psi_4^{44}$ obtained from the evolution of the Proca star discussed in Appendix~\ref{app:NumSetupNon}. From Fig.~\ref{fig:gw_upper_bound}, we conclude first that the extracted scalar $\Psi_4^{44}$ is consistent with truncation error of our numerical methods in \textit{both} the unperturbed and perturbed cases. Furthermore, we make sure that frequencies with $\omega_{44}M\lesssim 0.5$ are well-resolved by our numerical methods. Therefore, the upper bound of the emitted gravitational wave amplitude is $|r\Psi_4^{44}M|\lesssim 2\times 10^{-8}$. Since most of this residual is contained in frequencies above $\omega_{\rm GW}$ (see bottom panel of Fig.~\ref{fig:gw_upper_bound}), this is a conservative estimate. This implies the upper bound for the total emitted gravitational wave energy flux (in the $(\ell,m)=(4,4)$ mode and at frequency $|\omega_{\rm GW}M|=0.11$) is $\dot{E}_{\rm GW}^{44}\lesssim 2\times 10^{-16}$ at the reference probe field amplitude of $|J_\phi/J|\approx 2.9 \times 10^{-5}$. At this amplitude, the scalar unstable state exhibits an energy flux of $\dot{E}_\phi\approx 2\times 10^{-15}$. The two fluxes scale differently with the total angular momentum contained in the mode $J_\phi$: $\dot{E}_{\rm GW}\sim J_\phi^2$ and $\dot{E}_\phi\sim J_\phi$. This should be compared against the total available angular momentum that this mode can extract. In the absence of a fully nonlinear calculation, we estimate the latter as follows: the difference in angular momentum of the PS solution considered here (and in the main text), $J$, and the critical solution, $J_c$, at which the $m=2$ scalar state becomes unstable (as estimated by the zero-crossing of the extrapolation in Fig.~\ref{fig:testfieldcomp_ps}). This difference is $(J-J_c)/J\approx 5.2\times 10^{-4}=2J^{\rm strong}_\Phi$, which then implies \eqref{eq:gwupperbound} with the above upper bound.

\bibliography{bib.bib}

\end{document}